\documentclass[12pt,preprint]{aastex}

\usepackage{natbib}
\begin{document}

\title{2M1155--79 (= T Cha B): A Low-mass, Wide-separation Companion
  to the Nearby, ``Old'' T  Tauri Star T Cha} 

\author{J. H.\ Kastner\altaffilmark{1}, E. Thompson\altaffilmark{1,2},
  R. Montez Jr.\altaffilmark{1}, S.J. Murphy\altaffilmark{3},
  M.S. Bessell\altaffilmark{3}, G.G. Sacco\altaffilmark{1,4}}
\altaffiltext{1}{Center for Imaging Science and Laboratory for
  Multiwavelength Astrophysics, Rochester Institute of Technology, 54
  Lomb Memorial Drive, Rochester NY 14623 USA (jhk@cis.rit.edu)}
\altaffiltext{2}{West Irondequoit High School, Rochester, NY 14617}
\altaffiltext{3}{RSAA, Mount Stromlo Observatory, The Australian
  National University, ACT 2611, Australia}
\altaffiltext{4}{INAF-Osservatorio Astrofisico di Arcetri, Largo
  E. Fermi, 5, Firenze, 50125, Italy}

\begin{abstract}
  The early-K star T~Cha, a member of the relatively nearby
    ($D\approx 100$ pc) $\epsilon$ Cha Association, is 
  a relatively ``old'' (age $\sim$7 Myr) T Tauri star that is
  still sporadically accreting from an orbiting disk whose inner
  regions are evidently now being cleared by a close, substellar
  companion. We report the identification, via analysis of proper
  motions, serendipitous X-ray imaging spectroscopy, and followup
  optical spectroscopy, of a new member of the $\epsilon$ Cha
  Association that is very likely a low-mass companion to T~Cha at a
  projected separation of $\sim$38 kAU. The combined X-ray and optical
  spectroscopy data indicate that the companion, T Cha B (=
  2M1155$-$79), is a weak-lined T Tauri star (wTTS) of spectral type
  M3 and age $\stackrel{<}{\sim}$10 Myr.  The serendipitous X-ray
  (XMM-Newton) observation of T Cha B, which targeted T~Cha, also
  yields serendipitous detections of two background wTTS in the
  Chamaeleon cloud complex, including one newly discovered, low-mass
  member of the Cha cloud pre-MS population.  T~Cha becomes the third
  prominent example of a nearby, ``old'' yet still actively
  accreting, K-type pre-MS star/disk system (the others being TW Hya
  and V4046 Sgr) to feature a low-mass companion at very large
  (12--40 kAU) separation, suggesting that such wide-separation
  companions may affect the conditions and timescales for planet
  formation around solar-mass stars.
\end{abstract}

\section{Introduction}

The majority of stars likely form in binary systems
\citep[e.g.,][]{1994ARA&A..32..465M,2002ARA&A..40..349T}.  Given the
recent, rapid improvement in our knowledge of the demographics of
nearby ($D \stackrel{<}{\sim} 100$ pc), young (age $\sim10$--30 Myr)
stellar groups \citep[e.g.,][and references
therein]{2008hsf2.book..757T,2011ApJ...732...61Z}, we now have the opportunity to study,
at close range, the composition and, perhaps, dissolution of such
pre-main sequence (pre-MS) binaries during or just after the epoch of
planet formation. Indeed, initial studies of nearby, young
binaries make apparent that binary star
interactions could have profound consequences on planet-forming
circumstellar environments
\citep[e.g.,][]{2001ApJ...549..590P,2010ApJ...710..462A,2011ApJ...740L..17K}.

The $\epsilon$ Cha Association member T Cha \citep[spectral type K0, $D=109$ pc,
age $\sim7$ Myr;][]{2008hsf2.book..757T} is a rare example of a
nearby pre-MS star that displays evidence for
ongoing accretion \citep[other prominent examples 
are TW Hya, V4046 Sgr, and MP
Mus;][]{2008hsf2.book..757T}. The accretion rate of T~Cha is evidently
highly variable \citep[and is often undetectably small;][]{2008hsf2.book..757T,2009A&A...501.1013S}. Like
TW Hya and V4046 Sgr, T~Cha is a prototypical example of a
``transition disk'' object, i.e., a low-mass, pre-MS star orbited by a
relatively massive, dusty 
disk with a large (AU to tens of AU scale)  inner hole or gap. Such disk structure is possibly
indicative of the presence of young, massive planets
\citep[e.g.,][]{2011ApJ...738..131D}. Indeed, \citet{2011A&A...528L...7H}
recently obtained direct evidence (via aperture-masked adaptive optics
imaging) for a substellar object orbiting within the inner gap of the
T~Cha disk.  

Following the identification of apparent companions to HR 4796 and
V4046 Sgr at projected separations of $\sim$13.5 and $\sim$12.4 kAU, respectively
\citep{2008A&A...491..829K,2011ApJ...740L..17K}, we are searching for distant companions
associated with other pre-MS stars within $\sim100$ pc of Earth,
particularly nearby pre-MS stars that
are still orbited by and accreting from gaseous, circumstellar
disks. In this
paper, we report the identification, via analysis of catalogued proper motions,
serendipitous X-ray imaging spectroscopy, and followup optical
spectroscopy, of a new, M-type member of the $\epsilon$ 
Cha Association that is most likely a distant
  (projected separation $\sim38$ kAU) companion to T~Cha.

\section{Data and Results}

\subsection{Proper motions of red 2MASS stars near T Cha}

To search for candidate comoving, wide-separation companions to T~Cha,
we made use of the
VizieR\footnote{http://vizier.u-strasbg.fr/viz-bin/VizieR} table
browsing and source matching utilities available in
TOPCAT\footnote{Tool for OPerations on Catalogues And Tables;
  http://www.star.bris.ac.uk/~mbt/topcat/}.  We first compiled a list
of all 2MASS\footnote{The Two Micron All Sky Survey is a joint project
  of the University of Massachusetts and the Infrared Processing and
  Analysis Center/California Institute of Technology and is funded by
  NASA and the National Science Foundation.} sources within $20'$ of
T~Cha that have $K$ magnitudes and $J-K$ colors within the range
expected for low-mass (late K or M-type) stars within $\sim100$ pc,
i.e., $K \le 12$ and $0.8 \le J-K \le 1.2$. We then used the UCAC3
\citep{2010AJ....139.2184Z} proper motion catalog to select the subset
of these 2MASS sources whose measured proper motions (PMs) lie within
$\pm$15 mas yr$^{-1}$ of each of the PM components of T~Cha. 
This resulted in a list of 6 comoving companion candidates
to T~Cha.  Of these candidates, the closest PM match to T~Cha is 2MASS
11550485$-$7919108 (hereafter 2M1155$-$79); its UCAC3 PM
is indistinguishable from that of T Cha,
given the respective measurement uncertainties (Table~\ref{tbl:VJHKXMM}).  With an angular
separation of 6.36$'$, this candidate is also the closest in proximity
to T~Cha. 
The other five objects with PMs similar to T~Cha (and
  2M1155$-$79) all have RA PM components $|\mu_\alpha| \le 32$ mas
  yr$^{-1}$; these are most likely field stars unrelated to either the
  $\epsilon$ Cha Association or the background Chamaeleon cloud T
  Tauri star population \citep[given that three of the five are
  undetected in X-rays; see \S 2.2 and][]{1999A&A...341L..79T}.

\subsection{XMM-Newton archival data}

\begin{figure}
  \centering
\includegraphics[width=4in,angle=0]{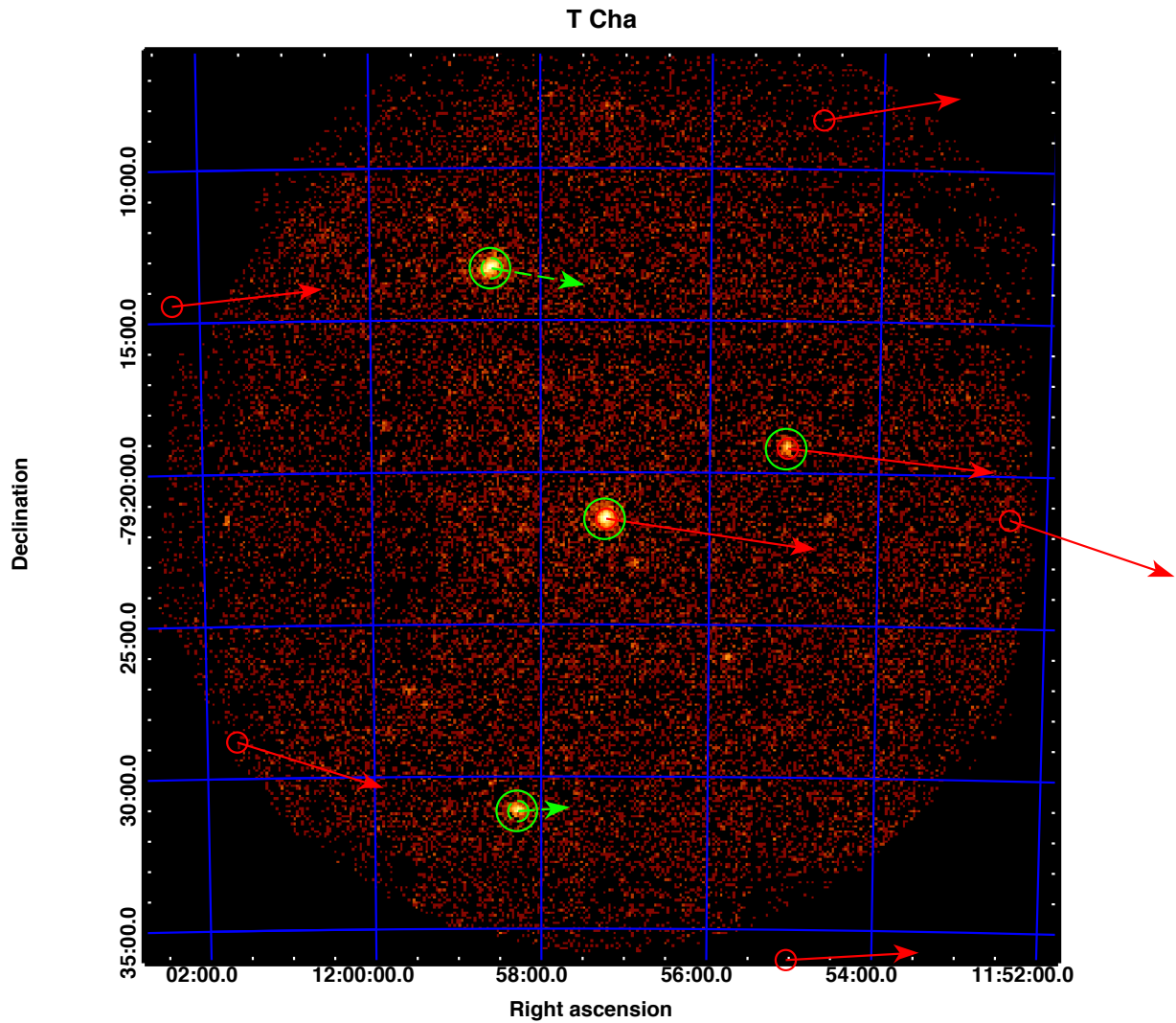}
\vspace{2.5in}
\caption{XMM/EPIC (0.2--12 keV) X-ray image centered on T Cha; N is up
  and E is to the left, and the imaged field is approximately $36'$ on
  a side. Green circles indicate the four brightest X-ray sources in
  the field: T~Cha (center) and (clockwise from upper left) RXJ1158.5,
  2M1155$-$79, and 2M1158$-$79. Positions and UCAC3 proper motions
  (PMs) of field stars with PMs similar to that of T Cha (see \S 2.1)
  are marked with red circles and vectors, respectively, where vectors
  represent angular displacements over a time of $10^4$ yr. The PMs of
 the background wTTS (RXJ1158.5 and 2M1158$-$79; see \S 2.2)
  are indicated as dashed green vectors.}
  \label{fig:XMMimage}
\end{figure}

Low-mass pre-MS stars are luminous X-ray sources, with typical
  X-ray luminosities (relative to bolometric) in the range $L_X/L_{bol}
  \sim$10$^{-4}$--10$^{-3}$ \citep[e.g.,][]{1997Sci...277...67K}. Hence, to
  confirm that 2M1155--79 is young, we examined an XMM-Newton
  observation, obtained obtained on 16 March 2009 UT, that targeted (and
  detected) T~Cha \citep[][]{2010A&A...519A.113G}. The observation
duration was 11.5 ks; the effective exposure times with the European
Photon Imaging Camera (EPIC) pn, MOS1, and MOS2 detectors were 3.3,
10.6, and 10.2 ks, respectively (the useful pn exposure time being
limited due to intervals of high background). The merged XMM/EPIC
0.5--2.0 keV image is displayed in Fig.~\ref{fig:XMMimage}, overlaid
with the positions and PM vectors of T~Cha and the half-dozen bright,
red 2MASS stars that have similar UCAC3 PMs (selected as described in
\S 2.1).  Of these 7 stars with similar PMs, only T~Cha and
2M1155$-$79 are detected as X-ray sources (note, however, that two of these
PM-selected stars do not fall within the XMM field of view).

The two other luminous X-ray sources in the field, which have UCAC3
catalog proper motions that are less than half those of T~Cha and
2M1155$-$79, correspond to 2MASS 11583429$-$7913175 (= RXJ
1158.5$-$7913; hereafter RXJ1158.5) and 2MASS 11581646$-$7931082
(hereafter 2M1158$-$79).  The former is a weak-line T Tauri star
(wTTS) of spectral type K3, previously identified by
\citet{1995A&AS..114..109A} as being associated with the Chamaeleon
cloud complex. The Cha clouds and their associated TTS population lie
$\sim$160 pc from Earth \citep{1997A&A...327.1194W} --- i.e.,
  $\sim60$ pc behind the $\epsilon$ Cha Association --- consistent
  with the small PM of RXJ1158.5 relative to $\epsilon$~Cha member
  T~Cha \citep[][]{1999A&A...341L..79T}.  Meanwhile, 2M1158$-$79 has
no previous references listed in SIMBAD\footnote{The SIMBAD database
  (http://simbad.u-strasbg.fr/simbad/) is operated at CDS, Strasbourg,
  France.}. Given its JHK magnitudes and colors relative to those of
RXJ1158.5 (Table~\ref{tbl:VJHKXMM}), and the similarity of their PMs,
2M1158$-$79 is therefore most likely a previously unidentified,
lower-mass (early M) member of the Cha cloud wTTS population. 
  Proper motions are unavailable for potential optical counterparts to the
  other, weaker X-ray sources in Fig.~\ref{fig:XMMimage}. 

\begin{figure}
  \centering
\includegraphics[width=6in,angle=90]{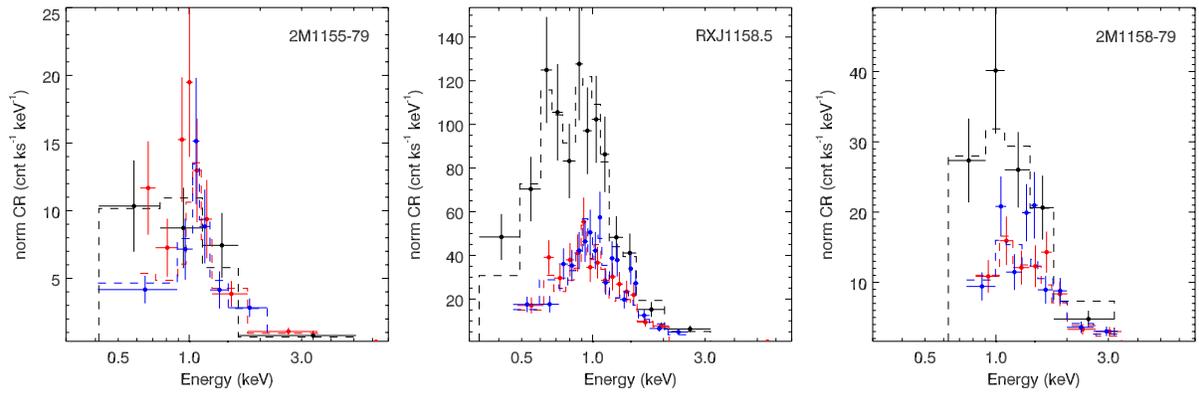}
\vspace{-1in}
\caption{XMM-Newton/EPIC spectra (pn: black points, MOS1: red points,
  MOS2: blue points) of 2M1155--79 and the two background (Cha cloud) wTTS
  in the T Cha field (RXJ1158.5 and 2M1158--79),
  overlaid with best-fit absorbed thermal plasma emission models
  (histograms, with colors corresponding to instrument data). }
  \label{fig:XMMspectra}
\end{figure}

We used the XMM Scientific Analysis System
(SAS\footnote{http://xmm.esa.int/sas} version 10.0.0) to extract pn,
MOS1, and MOS2 CCD spectra and responses for the three bright sources
that surround T~Cha in the EPIC image \citep[spectral fitting results
for T~Cha itself were presented
in][]{2010A&A...519A.113G}. Calibrations were performed using the
current calibration files (CCF) from release note 271,
21-Dec-2010. Background-subtracted total counts and count rates for
each source, as obtained from the pn, MOS1, and MOS2 spectral
extractions, are listed in Table~\ref{tbl:VJHKXMM} (the anomalously
low pn count rate of 2M1155--79 is the result of its position on a bad
row of that detector). We summarize the results of simultaneous fits of an absorbed
single-component thermal plasma emission model to the three (pn, MOS1,
MOS2) EPIC spectra of 2M1155$-$79 and 2M1158$-$79 and an absorbed
two-component plasma model to the EPIC spectra of RXJ1158.5 in
Table~\ref{tbl:VJHKXMM}, and we display the spectra overlaid with
these best-fit models in Fig.~\ref{fig:XMMspectra}  (a
second, lower-temperature 
component is necessary to model the low-energy regions of
the EPIC spectra of RXJ1158.5). The X-ray
properties of 2M1155--79 inferred from spectral fitting are similar to those of
the background (Cha cloud) wTTS RXJ1158.5 (and are typical of wTTS
more generally), consistent with a wTTS
classification for 2M1155--79 (\S 3.2). The best-fit model for
2M1158$-$79 (whose EPIC spectra are evidently somewhat harder than those
of 2M1155--79 and RXJ1158.5; Fig.~\ref{fig:XMMspectra}) indicates a
high plasma temperature and very large $L_X/L_{bol}$ ratio, suggesting
it may have been undergoing a strong flare during the short exposure
targeting T~Cha.

\subsection{Optical spectroscopy of T Cha and 2M1155--79}

\begin{figure}
  \centering
\includegraphics[width=2.4in,angle=90]{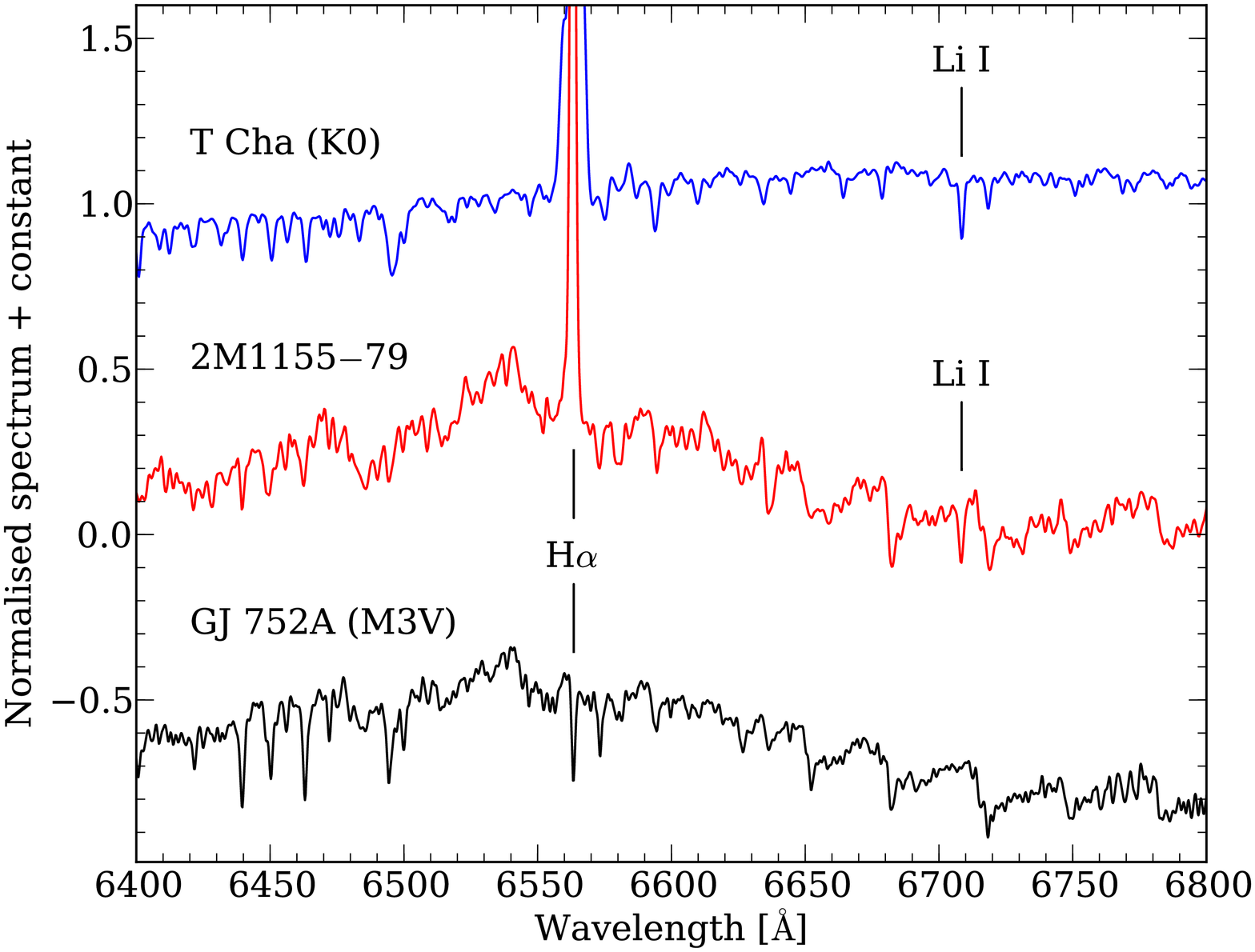}
\includegraphics[width=2.4in,angle=90]{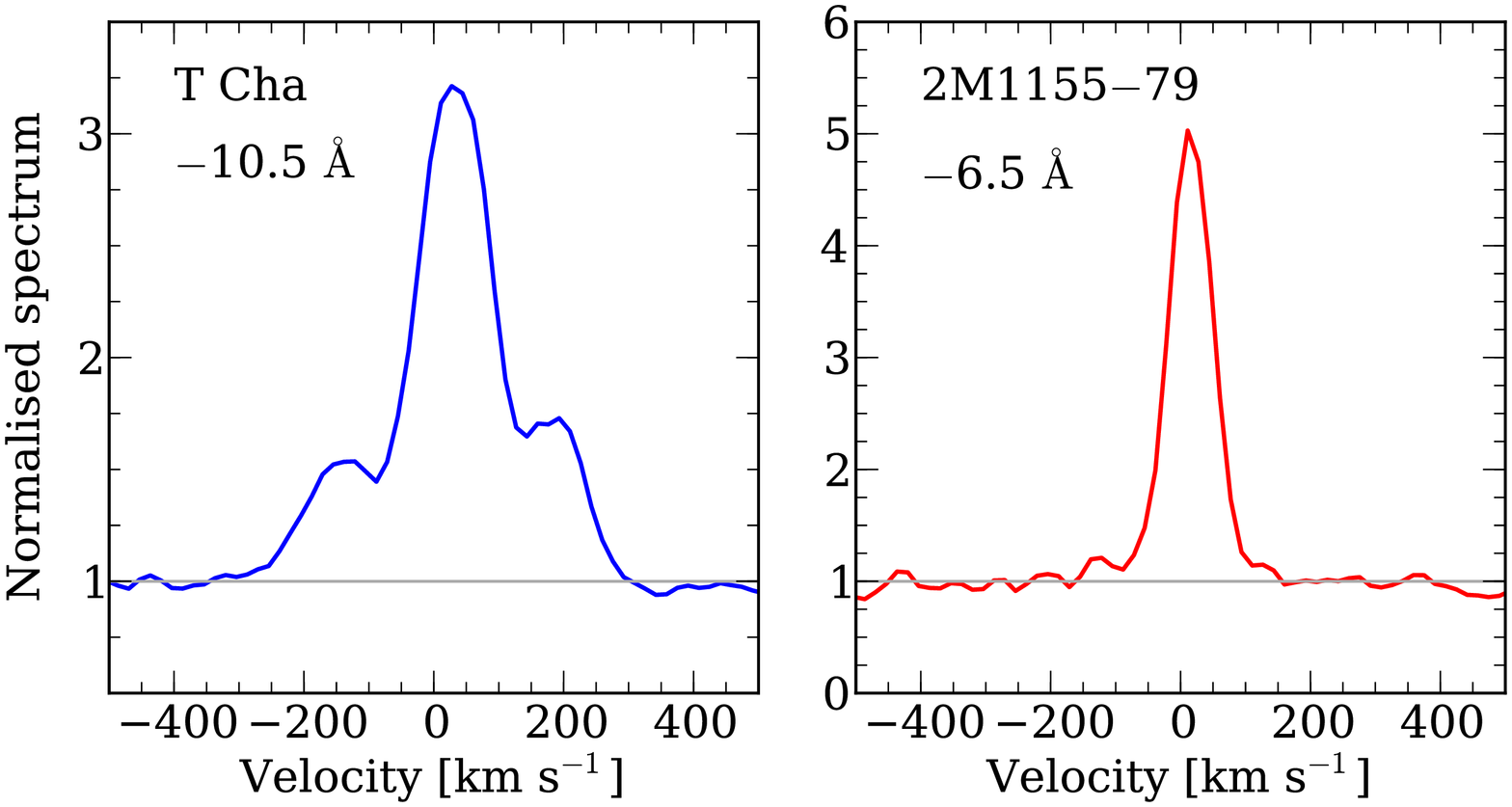}
\caption{{\it Left}: Siding Spring Observatory 2.3 m telescope 6400--6800 \AA\
   spectra of T Cha (blue), 2M1155--79 (red),
 and M3 field star GJ 752A (black) obtained with the WiFeS spectrograph. The
 positions of H$\alpha$ and the $\lambda$6708 Li {\sc i} line are
 indicated. {\it Center and right:} The H$\alpha$ emission line regions of the
  WiFeS spectra of T Cha and 2M1155--79 (= T Cha B). Line
  EWs are indicated in each frame.}
  \label{fig:spectra}
\end{figure}

Spectra of T~Cha and 2M1155--79 covering the H$\alpha$ and Li {\sc i}
$\lambda6708$ spectral region at resolution $\sim$7000 were obtained
with the Australian National University's Siding Spring Observatory
(SSO) 2.3 m telescope and WiFeS spectrometer
\citep{2007Ap&SS.310..255D} on 12 Oct.\ 2011
(Fig.~\ref{fig:spectra}). Exposure times were 2$\times$600 s and
  300 s for 2M1155--79 and T~Cha, respectively, and the stars were
  observed at airmass $\sim$2.5. Spectra were reduced as described in
  \citet{2011AJ....142..104R}; absolute flux calibration was not
  performed.  In the resulting, normalized spectra, both stars
display H$\alpha$ strongly in emission and show strong Li
$\lambda6708$ absorption lines (H$\alpha$ and Li $\lambda6708$ line
equivalent widths are reported in Table~\ref{tbl:VJHKXMM}).  The
photospheric absorption features in the spectrum of 2M1155--79 are
evidently a close match to those of the M3 field dwarf GJ 752A =
  HD 180617 \citep[][]{1994AJ....108.1437H}; see
  Fig.~\ref{fig:spectra} (left).
Hence, we adopt M3 as the spectral
  type of 2M1155--79 for purposes of the discussion in \S 3. 

\section{Discussion}

\subsection{2M1155--79 (= T Cha B): a likely wide-separation companion to T Cha}

Li is rapidly depleted in low-mass pre-MS stars, the $\lambda6708$
  photospheric absorption line becoming difficult
  to detect in nearby,  young
mid-M stars (even in high-resolution spectra) by the time such stars are of age $>10$ Myr
  \citep{2010ApJ...711..303Y}.  Hence, the detection of Li absorption
in the SSO WiFeS spectrum of 2M1155--79 (\S 2.3), combined with its
strong X-ray emission (\S 2.2) and H$\alpha$ emission (\S 2.2),
confirm that this star is young.  More specifically, the large
$\lambda$6708 Li {\sc i} EW we measure in the spectrum of 2M1155--79
is comparable to those of mid-M stars in the $\sim$8 Myr-old TW Hya
Association \citep{2011ApJ...727...62R} and 
  the $\sim$7 Myr-old $\epsilon$ Cha group \citep[even though
  2M1155--79 is among the lowest-mass $\epsilon$ Cha members
  known;][]{2008hsf2.book..757T}.  In contrast, its Li absorption line
  EW is a factor at least $\sim$3 larger than those of M stars (of all
  spectral subtype) in the $\sim$12 Myr-old $\beta$ Pic Moving Group
  \citep[][]{2010ApJ...711..303Y}. Hence its Li absorption line
  strength alone places an upper limit of $\sim$10 Myr on the age of
  2M1155--79.

\begin{figure}
  \centering
\includegraphics[width=5in,angle=0]{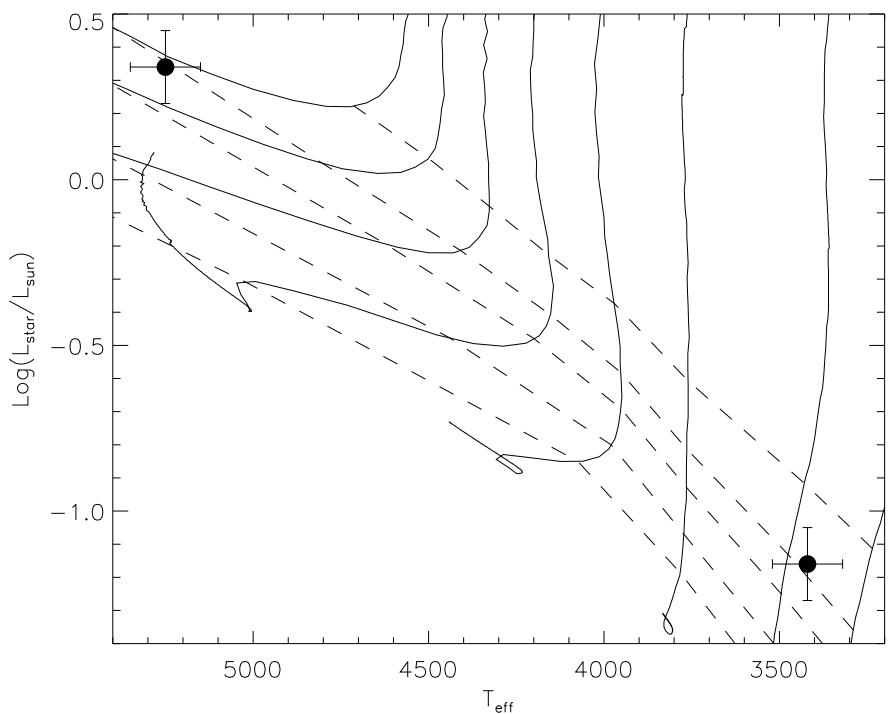}
\caption{The HR diagram positions of T Cha A (point at upper left) and
  2M1155--79 (= T Cha B; point at lower right) overlaid
  on pre-MS tracks from \citet[][]{2000A&A...358..593S}. 
  Error bars reflect estimated uncertainties of $\sim$25\% in luminosity
  (dominated by distance uncertainties) and $\sim$100 K in effective
  temperature. The evolutionary tracks
  (solid lines) correspond to masses of
0.2, 0.3, 0.5, 0.7, 0.9, 1.1, 1.3, and 1.5 $M_\odot$ (from lower right to upper left), while the
  isochrones (dashed lines) correspond to ages of 5, 10, 15, 25 and 40
  Myr (from upper right to lower left).}
  \label{fig:HRdiagram}
\end{figure}

  Adopting the data and methods in \citet[][]{1995ApJS..101..117K},
  the 2MASS magnitudes and M3 spectral type of 2M1155$-$79 suggest an
  effective temperature of 3400$\pm$100 K (where the error corresponds
  to an uncertainty of one spectral subclass) and --- assuming
  2M1155--79, like T~Cha, lies at $D=109$ pc
  \citep[][]{2008hsf2.book..757T} but, unlike T~Cha, suffers
  negligible reddening (given its low $N_H$; see below) --- a
  bolometric luminosity $\log{L_{bol}/L_\odot} = -1.16$. The pre-MS
  evolutionary tracks of \citet[][]{2000A&A...358..593S}, which are
  appropriate given the age and mass range of interest here
  \citep[e.g.,][]{2001ASPC..243..591L}, then indicate an age $\sim$10
  Myr and mass just under $\sim$0.3 $M_\odot$
  (Fig~\ref{fig:HRdiagram}). The former is consistent with the upper
  limit on the age of 2M1155--79 imposed by its strong Li absorption.
  In light of their similar proper motions \citep[\S 2.1; see also
  ][]{1999A&A...341L..79T} and the fact that (adopting $D=109$ pc) the
  two stars fall near the same (age 10 Myr) theoretical isochone in
  Fig~\ref{fig:HRdiagram} --- an isochrone that is, in turn, very
  similar to the estimated (7 Myr) age of the $\epsilon$ Cha group
  with which T Cha is kinematically associated \citep[][and references
  therein]{2008hsf2.book..757T} --- it appears that T~Cha and
  2M1155--79 are equidistant, coeval, and comoving. The latter star is
  therefore most likely a newly identified member of the $\epsilon$
  Cha Association.

  Furthermore, given the low surface density of known members of the
  $\epsilon$ Cha group \citep[$\sim0.18$
  degree$^{-2}$;][]{2008hsf2.book..757T}, the Poisson probability
  $P$ that T~Cha and 2M1155--79 represent a chance alignment of two
  (otherwise unrelated) $\epsilon$ Cha Association stars within the
  ($\sim0.2$ degree$^2$) XMM/EPIC field is vanishingly small ($P
  \approx 6\times10^{-4}$). We therefore assert that these two stars
  very likely comprise a very wide (projected separation $\sim$38 kAU)
  binary system; hence, hereafter, we refer to 2M1155--79 as T Cha B.

\subsection{The contrasting natures of T Cha A and B}

The H$\alpha$ emission line strength of T~Cha displayed in our
spectrum (Table~\ref{tbl:VJHKXMM} and Fig.~\ref{fig:spectra}, center) lies near the middle of the (wide)
range previously measured for this star, and is only marginally
consistent with ``classical'' (as opposed to weak-lined) T Tauri star
status \citep[the T~Cha H$\alpha$ line has
also occasionally been observed in 
absorption;][]{2008hsf2.book..757T,2009A&A...501.1013S}. The 
H$\alpha$ linewidth at 10\% of peak intensity at the time of our
observations, $W_{10} =$ 475 km s$^{-1}$, would place T~Cha
among the weakly accreting T Tauri stars --- with an inferred
accretion  rate of $dM/dt \sim3\times10^{-9}$ $M_\odot$ yr$^{-1}$ --- given
the empirical relationship between $dM/dt$ and $W_{10}$ described in
\citet{2004A&A...424..603N}. 

In contrast, we measure $W_{10} =$ 149 km s$^{-1}$ for T Cha B
(Fig.~\ref{fig:spectra}, right), suggesting $dM/dt < 10^{-11}$
$M_\odot$ yr$^{-1}$ --- consistent with the wTTS classification for
this star one would determine from its H$\alpha$ emission-line EW
\citep[Table~\ref{tbl:VJHKXMM}; e.g.,][]{2003AJ....126.2997B}.  
  Furthermore, whereas T~Cha A displays a strong mid- to far-infrared
  excess in 2MASS and Wide-field Infrared Survey Explorer\footnote{The
    Wide-field Infrared Survey Explorer is a joint project of the
    University of California, Los Angeles, and the Jet Propulsion
    Laboratory/California Institute of Technology, and is funded by
    NASA.} (WISE) data (e.g., $K-W3\sim 2.3$, compared with $K-W3\sim
  0.5$ for the wTTS RXJ1158.5; Table~\ref{tbl:VJHKXMM}) and is
  evidently subject to varying degrees of obscuration by its dusty
  circumstellar disk \citep[][]{2009A&A...501.1013S}, T~Cha B displays
  only a modest mid-infrared excess ($K-W3\sim 0.8$) --- although the star
  evidently has a substantial $\sim$20 $\mu$m flux excess (as does the background wTTS
  2M1158--79; Table~\ref{tbl:VJHKXMM}), suggesting a significant
  mass of cool circumstellar dust. The presence of disk gas
  around T Cha A --- and the apparent absence of such intervening disk
  gas, along the line of sight to T Cha B --- is also evident in the
  large contrast between their
  respective X-ray absorbing columns (Table~\ref{tbl:VJHKXMM}).

\section{Conclusions}

T~Cha is the third prominent example of a nearby
($D\stackrel{<}{\sim}100$ pc, ``old'' (age $\sim$ 10 Myr) yet still
actively accreting, K-type pre-MS star/disk system (the others being
TW Hya and V4046 Sgr) that is now known to feature  candidate
  low-mass companions at very large ($\sim$12--40 kAU)
separation. The projected separation of the apparent T Cha A/B
pair, $\sim$38 kAU, is very similar to that of the TW Hya Association
(TWA) brown dwarf candidate TWA 28 from TW Hya itself \citep[$\sim$41
kAU;][]{2008A&A...489..825T}, and is much larger than the projected
separation of V4046 Sgr AB and C[D] \citep[$\sim$12.4
kAU;][]{2011ApJ...740L..17K}. Other examples of nearby, $\sim$10
Myr-old binary systems with components at wide ($\sim$3--13 kAU)
separation that feature dusty circumstellar disks are TWA members HR
4796 \citep[TWA 11;][]{2008A&A...491..829K} and TWA 30 \citep[whose
two components also display evidence for ongoing
accretion;][]{2010AJ....140.1486L}.  The projected
separations of four of these systems (T Cha A and B; TW Hya and TWA
28; V4046 Sgr AB and C[D]; and HR 4796 AB and C) rival or exceed
those of the widest known main-sequence binaries in the solar
neighborhood \citep[$\sim$20 kAU;][]{1990AJ....100.1968C}. It may
  be that we are observing these young binary systems at a ``fragile''
  stage of their evolution, during which they are particularly
  susceptible to dissolution (e.g., via encounters with the older
  field star population).  Indeed, the large separation of T~Cha A
  and B implies an orbital period $\sim5$ Myr, similar to the age of
  the system itself. High resolution
  spectroscopy and precise proper motion determinations --- capable of
  establishing the relative space motion of the two stars to within
  $\sim0.1$ kms$^{-1}$ --- are required to ascertain whether they are
  in fact gravitationally bound.

Furthermore, depending on the nature of the substellar companion
orbiting within the inner hole of the T Cha disk
\citep{2011A&A...528L...7H}, T Cha --- like $\beta$ Pic Moving Group
member V4046 Sgr and TWA members HD 98800, Hen 3--600, and HR 4796 ---
might be considered a {\it hierarchical} binary system with long-lived
disk. The presence of wide-separation companions could be pointing to
the profound effects of companion-disk interactions, and perhaps hierarchical
binary dissolution, on the conditions and timescales for planet
formation in such systems \citep[][and references therein]{2011ApJ...740L..17K}. In
particular, given the advanced ages of these five disk-retaining
hierarchical binaries, one might speculate that, in each case, the
presence of a companion(s) has either inhibited
planet formation processes \citep[see also][]{2001ApJ...549..590P} or
extended the time available for such processes well beyond the
``nominal'' few Myr typically inferred for disk lifetimes
\citep[e.g.,][and references therein]{2009ApJ...698....1C}.

\acknowledgments{\it We thank the anonymous referee for helpful
  comments and suggestions. This research was supported by grants to RIT from the National
  Science Foundation (award AST--1108950) and the NASA
  Astrophysics Data Analysis Program (award NNX09AC96G).}


\begin{table}[ht]
\scriptsize
  \begin{center}
    \caption{\sc Data and Results for T Cha field XMM sources$^a$}
 \begin{tabular}{ccccc}
\hline
\hline
& T Cha & 2M1155-79 & RXJ1158.5-7913 & 2M1158--79\\
\hline
$\mu_\alpha$ (mas yr$^{-1}$) & $-41.2$ (3.2) & $-40.6$ (5.1) & $-18.3$ (3.8) & $-10.1$
(2.9) \\
$\mu_\delta$ (mas yr$^{-1}$) & $-6.1$ (1.4) & $-4.7$ (5.6) & $-3.1$ (1.9) & 0.7 (2.4) \\
$J$ (mag, 1.25 $\mu$m) & 8.955 (0.027) & 11.216 (0.022) & 9.745 (0.024) & 12.232 (0.026) \\
$H$ (mag, 1.65 $\mu$m) & 7.860 (0.046) & 10.465 (0.026) & 8.959 (0.024) & 11.544 (0.026) \\
$K$ (mag, 2.2 $\mu$m) & 6.954 (0.018) & 10.079  (0.021) & 8.653 (0.019) & 11.291 (0.024) \\
$W1$ (mag, 3.4 $\mu$m) & 5.899  (0.047) & 9.894  (0.024) &8.497 (0.024) & 11.144 (0.025) \\ 
$W2$ (mag, 4.6 $\mu$m) & 4.974 (0.030) & 9.667    (0.021) &8.384 (0.021) & 11.153(0.022) \\
$W3$ (mag, 12 $\mu$m) & 4.631  (0.017) & 9.273   (0.033) &8.188 (0.024) & 10.974	(0.092) \\
$W4$ (mag, 22 $\mu$m) & 2.601 (0.017) & 7.195   (0.081) &7.864 (0.132) & 8.683 (0.239) \\
EW, H$\alpha$ (\AA)       & $-10.5$ (0.4) & $-6.5$ (0.5) & $-3.1^b$ & ... \\
EW, Li $\lambda6708$ (m\AA) & 350 (50) & 550 (100) & 600$^b$ & ... \\
EPIC Counts$^c$ & 432, 660, 687 & 48, 148, 122 &  363, 441, 513  & 150, 276, 303
\\   
Count Rate$^c$ (cnt ks$^{-1}$) & 132, 63, 61 & 13, 13, 9.3 & 110, 41, 45 & 43, 24, 25 \\ 
$N_{H}$ ($\times10^{21}$ cm$^{-2}$) & 9.7$^d$ & 0.16 (0--0.79) & 0.47
(0.30--0.68) & 1.5 (0.6--2.4) 
\\ 
$kT_X$ (keV) & ... & 2.1 (1.7--2.8) & 0.65 (0.60--0.74), 2.3 (2.1--2.7) &
4.7 (3.4--7.6) \\ 
$F_{X}$$^e$ (erg cm$^{-2}$ s$^{-1}$) & $2.0\times10^{-12}$$^d$ & $1.2\times10^{-13}$ &
$5.9\times10^{-13}$ & $5.9\times10^{-13}$ \\
$\log{(L_X/L_{bol})}^f$ & $-3.91^d$ & $-3.19$ & $-3.11$ & $-2.1$: \\
\hline
  \end{tabular}
\label{tbl:VJHKXMM}
\end{center}
NOTES:\\
a) Obtained via the observations and analysis
described in text (\S\S 2, 3) unless otherwise indicated.\\
b) EW data from \citet{1995A&AS..114..109A}.\\
c) pn, MOS1, and MOS2 total counts and net (background-subtracted)
count rates. \\
d) Based on X-ray spectral fit results listed in \citet{2010A&A...519A.113G}. \\
e) Intrinsic (``unabsorbed'') X-ray flux.\\
f) Bolometric fluxes for RXJ1158.5 and 2M1158--79 were
estimated as described in \S 3.1 (for 2M1155--79). Both the bolometric
flux and steady-state X-ray flux of 2M1158--79 are uncertain.
\end{table}

\newpage

\end{document}